\documentclass[10pt,journal,letterpaper]{IEEEtran}
\IEEEoverridecommandlockouts
\usepackage[T1]{fontenc}
\usepackage{orcidlink}
\usepackage{setspace}

\IEEEoverridecommandlockouts
\usepackage{fancyhdr}
\usepackage[numbers, sort&compress]{natbib}

\usepackage{float}
\usepackage{enumerate}
\usepackage{hyperref}
\usepackage{amsmath,amssymb,amsfonts}
\usepackage{algorithmic}
\usepackage[ruled,vlined]{algorithm2e}
\usepackage{graphicx}
\usepackage{textcomp}
\usepackage{xcolor}
\usepackage{physics}
\usepackage[font=footnotesize, skip=0pt]{caption}
\usepackage{subcaption}
\usepackage{comment}

\usepackage{booktabs}
\usepackage{multirow}
\def\BibTeX{{\rm B\kern-.05em{\sc i\kern-.025em b}\kern-.08em
    T\kern-.1667em\lower.7ex\hbox{E}\kern-.125emX}}
\begin{document}

\title{Obfuscated Quantum and Post-Quantum Cryptography}
 
\author{Anju~Rani$^{*}$, Xiaoyu~Ai$^{*}$, Aman Gupta$^{*}$, Ravi Singh Adhikari$^{*}$ and~Robert~Malaney$^*$ 
\thanks{$^*$Anju Rani, Xiaoyu Ai, Aman Gupta, Ravi Singh Adhikari, and Robert Malaney are with the School of Electrical Engineering and Telecommunications, University of New South Wales, Sydney, Australia.}
}

\maketitle
\begin{abstract}
In this work, we present an experimental deployment of a new design for combined quantum key distribution (QKD) and post-quantum cryptography (PQC). Novel to our system is the dynamic obfuscation of the QKD-PQC sequence of operations, the number of operations, and parameters related to the operations; coupled to the integration of a GPS-free quantum synchronization protocol within the QKD process. We compare the performance and overhead of our QKD-PQC system relative to a standard QKD system with one-time pad encryption, demonstrating that our design can operate in real time with little additional overhead caused by the new security features. Since our system can offer additional defensive strategies against a wide spectrum of practical attacks that undermine deployed QKD, PQC, and certain combinations of these two primitives, we suggest that our design represents one of the most secure communication systems currently available. Given the dynamic nature of its obfuscation attributes, our new system can also be adapted in the field to defeat yet-to-be-discovered practical attacks.
\end{abstract}



\setlength{\parindent}{4ex} 
\setlength{\parskip}{1ex}
\renewcommand{\baselinestretch}{1.5} 

\patchcmd{\thebibliography}{\section*{\refname}}{}{}{}

\section{Introduction}
The advancement of quantum computing poses a significant threat to the classical cryptographic algorithms widely used today. Algorithms based on the integer factorization problem, the discrete logarithm problem, and the elliptic-curve discrete logarithm problem are all vulnerable to a quantum adversary (someone equipped with a sufficiently powerful quantum computer) and can be broken in polynomial time~\cite{wang2024first}. Consequently, cryptographic solutions resistant to quantum algorithms are essential, not only to safeguard future communication infrastructures but also to ensure secure key distribution and data protection across emerging quantum networks~\cite{liu2024post}.
The two primary strategies that have been proposed to address these challenges are quantum key distribution (QKD) and post-quantum cryptography (PQC).

\par QKD, in principle, enables information-theoretic secure communication between two legitimate parties even in the presence of an adversarial eavesdropper. Unlike classical cryptography methods that rely on computational hardness of the relevant mathematical problems, QKD leverages inherent quantum features, such as the no-cloning theorem, indistinguishability of non-orthogonal quantum states, and quantum entanglement, to achieve information-theoretic security, offering provable guarantees that remain valid regardless of an adversary's computational power.  QKD has achieved significant milestones in global deployment, including high secret key generation rates~\cite{islam2017provably,yuan201810}, long-distance quantum communication~\cite{pittaluga2021600,xie2022breaking}, photonic integration~\cite{sibson2017chip,bunandar2018metropolitan}, and the development of terrestrial and satellite-based quantum networks~\cite{chen2021integrated}. Since the implementation of the original prepare-and-measure QKD protocol~\cite{bennett2014quantum} and the entanglement-based protocol~\cite{ekert1991quantum},  many new QKD protocols have been proposed (for a review, see~\cite{kumar2025brief}). However, the \textit{practical} implementation of any QKD protocol continues to face ongoing debates in terms of the real-world security it currently provides~\cite{NSA,renner2023quantum}. These debates generally focus on the fact that any implementation of QKD can generate imperfections and loopholes, due to deviations from the mathematical assumptions underpinning the chosen protocol~\cite{tupkary2025qkd,huang2018implementation,rani2023experimental}. It is important to recall that a prerequisite for all QKD protocols is the presence of a pre-shared key (PSK) known only to the sender-receiver pair (the reason why QKD is termed a \textit{key-growing} protocol).

\par The alternative to QKD, namely PQC,  encompasses various cryptographic families, including lattice-based, hash-based, code-based, multivariate, and isogeny schemes. PQC does not provide information-theoretic security, but remains, as far as we currently know, computationally secure against a quantum adversary; a form of security we henceforth refer to as quantum resistant security. Assuming that a shared symmetric secret key is available of at least 256 bits length, the Advanced Encryption Standard (AES) is also known to be quantum resistant~\cite{bonnetain2019quantum}. However, there are still vulnerabilities.
For example, the most modern encryption schemes used in transport layer security (TLS) v1.3, AES-Galois Counter Mode and AES-Counter have been shown to be vulnerable in implementation~\cite{bock2016nonce,mattsson2024collision,chung2025making,albertini2022abuse,kampanakis2023practical,jayasinghe2014advanced,tienteu2023template}.
Concerning scenarios where no such symmetric key is \textit{a priori} available, multiple protocols are being investigated.
The National Institute of Standards  and Technology (NIST) has announced the candidate lists for the key exchange and digital signature schemes for standardization~\cite{moody2021nist}. Among the five selected schemes, CRYSTALS-Kyber~\cite{bos2018crystals} and HQC~\cite{melchor2018hamming} are selected as key encapsulation schemes, and CRYSTALS Dilithium~\cite{ducas2018crystals}, FALCON~\cite{fouque2018falcon}, and SPHINCS+~\cite{bernstein2019sphincs+} are the three selected digital signature schemes. 
Similarly to QKD and AES, the implementation of any PQC protocol will always give rise to the possibility of attack avenues. 
For more details on the status of PQC, the challenges faced, and the threats in implementation, the reader is referred to~\cite{PQC12, nguyen2025security,alnahawi2025sok}.

\begin{figure*}[htp]
    \centering
    \includegraphics[width=1\linewidth]{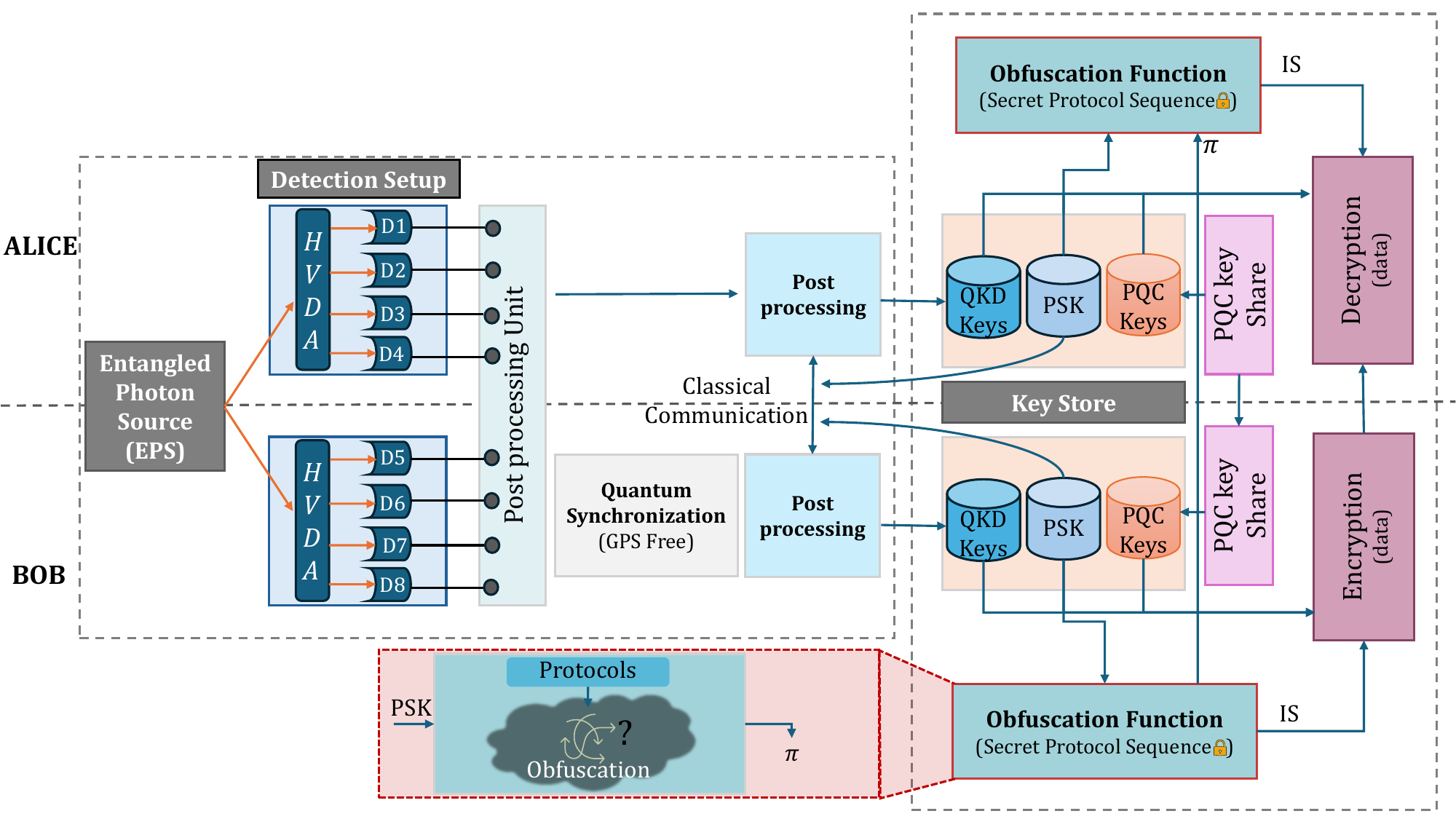}
    \caption{The experimental setup and the combined QKD-PQC encryption–decryption module. This includes the optical setup with single photon detectors D1 through D4 at Alice detecting the horizontal ($H$), vertical ($V$),  diagonal ($D$), or anti-diagonal ($A$) basis states, respectively (D5 through D8 similarly for Bob); and the post-processing unit, followed by quantum synchronization embedded within the QKD process. A software-implemented obfuscation function is used to determine a secret mode of system operations and settings, provided as an instruction sequence (IS), which is encrypted to form $\pi$ (an encrypted message (or identifier) sent from Bob to Alice so that the IS is known only to them). An example IS could be the type and order of the encryption-decryption protocols. Critical to the experiment is the pre-shared key (PSK), assumed to be known only to the data sender (Bob) and data receiver (Alice) and  \textit{a priori} shared between them. Part of the PSK is required for QKD, part for our obfuscation function, and part for AES. The EPS is external to both Alice and Bob. The bottom centered box (pink)  illustrates the internal operations carried out within the obfuscation function module (bottom-blue box).}
    \label{fig:Esetup}
\end{figure*} 

\par Considering the practical limitations perceived in both QKD and PQC, several hybrid QKD-PQC\footnote{A note on terminology: The term `QKD-PQC' will henceforth be used in this work to cover combinations of QKD, coupled to one or more quantum resistant algorithms. Quantum resistant AES is not formally classified as a member of the PQC family. However, we will ignore such definitional formalities in our use of the term `QKD-PQC'.} schemes have emerged in recent years. 
Some works combine QKD and PQC keys into a single new key ~\cite{ricci2024hybrid,zeng2024practical,bindel2019hybrid,garms2024experimental,dowling2020many,bruckner2023end,10648124,buruaga2025hybrid,garcia2023quantum,blanco2025hybrid}, some use PQC to authenticate or establish trust within QKD~\cite{geitz2023hybrid,ghashghaei2024enhancing,wang2021experimental,prakasan2022authenticated,marchsreiter2023pqc}, and some use PQC encryption for security coupled with QKD encryption for route anonymity~\cite{otero2025onion}. Other works use PQC-encrypted reconciliation information to improve the secret key rates of QKD~\cite{djordjevic2020joint}, use QKD as a seed generator to improve the security of PQC~\cite{djordjevic2021qkd},  discuss how
quantum random number generators can improve the combination of QKD and PQC~\cite{azarderakhsh2025integrating}, or consider the advantages of distributed architectures to the QKD-PQC paradigm~\cite{kim1}.
These hybrid schemes offer the potential advantage of QKD's information-theoretic security alongside the quantum resistant security of the PQC schemes - and point to a paradigm for real-world communication security in the quantum computing era. If a successful attack on one primitive is available (for example, due to a flawed implementation),  the other primitive still provides its inherent security. 
\textit{In this work, we take this paradigm one step further and introduce an information-theoretic secure obfuscation\footnote{Note, we adopt here a plain dictionary meaning of this word - the action of making something hidden, or unintelligible. Our use of "obfuscation" \textit{does not} have any other specific meaning provided to it by other works, such as its use in quantum circuit obfuscation.} of the system-level configuration setting that instructs how to combine the QKD and PQC primitives.} 
One key outcome of this is that only the data sender and the data receiver are aware of the sequence and number of the QKD and PQC implementations, as well as the important parameters of each primitive. Although this new obfuscation is redundant to any user when QKD is perfectly implemented and redundant to a classical-only user when PQC is perfectly implemented, the additional layer of security allows for the foiling of known practical attacks jointly on each primitive in addition to enhanced protection against yet-to-be-discovered practical attacks on each primitive. 

\par The work presented here builds on our previous experiments~\cite{rani2025combined} which demonstrated a QKD-PQC implementation that sequentially combines a QKD primitive based on BBM92 with quantum resistant AES (with a 256-bit key). Here, we extend that work by not only introducing the aforementioned obfuscation of the QKD-PQC configuration setup, but also the integration of a self-synchronizing mechanism in our QKD key generation system that eliminates the dependence on the global positioning system (GPS) or other classical synchronization methods. Our synchronization protocol utilizes the tight temporal correlation (at birth) of pairs of polarization-entangled photons. This provides for an additional level of time-based security against an intercept-resend attack, albeit one that is not information-theoretic secure (see the later discussion).
Also, by integrating time synchronization and QKD, we can significantly reduce hardware overhead.
The new integrated QKD-PQC and quantum synchronized system that we implement is shown in Fig.~\ref{fig:Esetup}. We suggest that this system represents the most secure practical communication system currently available.

\par The remainder of this paper is structured as follows. Sec.~\ref{sec:exp} provides the experimental implementation of the QKD protocol, including the details of our synchronization algorithm.  Sec.~\ref{sec: combined system}  explains the design of our combined quantum and post-quantum system, and Sec.~\ref{sec:results} presents the results, with a focus on the penalty in time incurred by the inclusion of QKD-PQC obfuscation. Finally, Sec.~\ref{sec:conclusions} concludes with some remarks on our findings.

\section{\label{sec:Bg}System Model}

\textcolor{black}{As described above, our system is useful \textit{relative to existing QKD-PQC architectures}, only on the premise that the normal information-theoretic security of the QKD keys is not applicable and that the quantum-resistant security of a single-use PQC primitive is not applicable. As such, our additional security is not formally defined, and is best described as an additional resource challenge to the adversary. For example, in the presence of two side-channel attacks that render our premise valid, the adversary is challenged on two new fronts: (i) She must consider all possible combinations of parameters and sequence primitives allowed by the system; and then (ii) find a series of side-channel attacks that would be successful against a significant fraction of those combinations (on average 1/2). The only portion of our system under our attack premise that can be information-theoretic secure is the security of the instruction sequence (the combinations) itself.}

\subsection{\label{sec:exp}QKD Implementation}
We implement an entangled version of QKD, namely BBM92~\cite{mishra2022bbm92}. In entanglement-based QKD, the source can be an intermediary device that distributes highly correlated photons to two separate parties, Alice and Bob, via two independent quantum channels, which in our setup are largely free-space channels. Our prototype QKD system is eventually aimed at low Earth orbit deployment, in which case the entangled photon source (EPS) will be on board a satellite communicating to two widely separated ground stations. As a consequence of the tight temporal correlation at pair production, the EPS offers additional advantages to our QKD system through a GPS-free synchronization protocol,  something of importance for satellite systems. Therefore, unless otherwise stated, in the following we assume that the EPS is embedded on a satellite (or other detached third-party system communicating through free space). For fiber-based QKD implementations, the reader is referred to one of the many papers in this area \cite{qkdJE}. For fiber-based time synchronization, see~\cite{synJE}.

\par Our EPS~(fiber-coupled prior to launch into free space) is used to generate polarization-entangled photon pairs, known as the signal and idler. The state produced by our source is given by
$\ket{\psi} = \frac{1}{\sqrt 2}(\ket{H}_{sig}\ket{V}_{id}+\ket{V}_{sig}\ket{H}_{id})$,
where $\ket{H}_{sig}$ and $\ket{V}_{sig}$ represent horizontally and vertically polarized photons, respectively; and the subscripts $sig$ and $id$ indicate the signal and idler, respectively. The photon pairs co-propagate in free space and are separated and distributed to Alice and Bob. 

\par \textcolor{black}{The received photons at Alice and Bob are measured by projecting their polarization states onto four basis states, $\{H, V, D, A\}$, using four single-photon detectors. Here,  $H, V, D$, and $A$ represent the horizontal, vertical, diagonal, and anti-diagonal  polarization basis states, respectively.
The detected photons and their arrival times are recorded for post-processing and subsequent generation of the QKD keys.}
The post-processing hardware unit consists of a time tagging device with an inbuilt field programmable gate array (FPGA) unit, as shown in Fig.~\ref{fig:Esetup}. 
Alice and Bob (two personal computers) also perform post-processing steps, 
including interactive time domain filtering, sifting, QBER estimation, error correction, and privacy amplification.  
More details on the post-processing follow, in which a QKD `session' is defined as over by the completion of the privacy amplification described below.

\subsubsection*{\label{sec:bitmap}Bit Mapping and Sifting}
Once signal acquisition is completed, the arrival times of photons registered from each detector are placed in time tag arrays $t_a$ and $t_b$ at Alice and Bob, respectively. This information must first be corrected for any synchronization errors (see~\ref{sec:timetag}).
\textcolor{black}{Once the time tag arrays are synchronized, Alice determines two bits for each time tag in $t_a$ that indicate which of her four detectors registers the pulse (similar for Bob). }
The first mapped bit (corresponding to $H$, $V$, $D$, $A$) is then stored in a bit array, $k_a$ ($k_b$ for Bob), and the second bit (corresponding to the basis used) is stored in an array, $b_a$ ($b_b$ for Bob). \textcolor{black}{Let the length of $k_a$ and $k_b$ be ${N_{\text{raw}}}$ at this point.} 
To perform the basis sifting, Alice and Bob exchange their $b_a$ and $b_b$ via classical communications - only bits measured in the same basis are retained. Each step performed through classical communication undergoes encryption as described by an instruction sequence (discussed later).
The length of the remaining $k_a$ and $k_b$ will be \textcolor{black}{on average}, ${N_{\text{raw}}}/{2}$.

\subsubsection*{QBER Estimation} \textcolor{black}{In this step, Alice randomly selects a subset (with the size of ${N_{\text{raw}}}/{4}$) of bits from $k_a$ and sends them to Bob, who then compares these received bits with his corresponding bits in $k_b$ and estimates the QBER (denoted by $Q$).} 
Next, Bob discards the bits used in the comparison and informs Alice of the estimated QBER. Alice then also removes the selected bits from $k_a$. We denote the length of $k_a$ and $k_b$ by the end of this step as $N$.

\subsubsection*{Reconciliation with Message Authentication Codes (MAC)}
The bit vectors $k_a$ and $k_b$ are generally not identical due to quantum channel noise, time-tag errors, and synchronization errors. To reconcile the keys, a syndrome decoding technique based on an appropriate error correction code is adopted.   Authentication in reconciliation is achieved through a Wegman-Carter MAC (WC-MAC)~\cite{wegman1981new}, which is an information-theoretic many-time-MAC scheme.
We note that the advantage of using this MAC in QKD is that it consumes only $\log_2(l^{s_b})$ bit keys from the previous QKD session, where $l^{s_b}$ is the length of the syndrome, $s_b$. 

\subsubsection*{Privacy Amplification}
Assuming a successful reconciliation, Alice and Bob will have two identical bit vectors.
In the privacy amplification step, they  extract QKD keys from these bit vectors. Specifically, Alice first computes the QKD key rate, $r$, in bits per pulse, using the asymptotic analysis
of the BBM92 protocol~\cite{ma2007quantum,cai2009finite}:  
$r = 1-h(Q) - L$,
where $h$ is the binary entropy function, $L=1-R_c$. 
is the ratio of the information disclosed to Eve  during the error correction step to the length $N$.
\textcolor{black}{Note, $R_c=(N-l^{s_b})/N = 1-f_Eh(Q)$, where $f_E=l^{s_b}/(Nh(Q))$ is the reconciliation efficiency of the code.} Then, Bob will follow the instructions described in~\cite{bourgoin2015experimental} to construct a 2-universal hashing matrix, $\mathbf{T}$, with the dimension of $\lceil rN\rceil$-by-$N$ and send $\mathbf{T}$ to Alice. Finally, both parties generate their respective secret keys by multiplying their reconciled bit vectors with the matrix $\mathbf{T}$.

\subsection{\label{sec:timetag}QKD Synchronization} 
In QKD protocols, precise time synchronization improves secure key generation by improving the signal-to-noise ratio, thus lowering the QBER~\cite{pljonkin2017synchronization,pelet2023operational,basso2021quantum,villasenor2024towards,pelet2025entanglement,lafler2023quantum,spiess2023clock}. 
As such, we introduce such a protocol into our system here. Our synchronization protocol, which comprises combining algorithms~1 and 2 in the supplementary file~\cite{alganju}, estimates a relative time offset ($\tau$) and time drift ($\tau_{drift}$) between Alice and Bob. This must be done prior to the commencement of QKD, otherwise the QKD will likely fail.
Bob performs a two-step synchronization procedure between Alice’s time tag array $t_a$ and Bob's time tag array $t_b$. To achieve this, these time tag arrays are manipulated sequentially into a series of significantly shorter time tag arrays $\hat{t}_a$ and $\hat{t}_b$ for Alice and Bob, respectively. Each shorter time tag array is associated with a `round,' a number which specifies the creation order (first tag, round 1: second tag, round 2, etc.); and spans an acquisition period $T_{acq}$, meaning that a time difference between its first and last time tag is approximately $T_{acq}$.  Note that to protect against a time tag attack~\cite{pahali2022cryptographic}, Alice encrypts her time tag information using some specific instruction sequence to form $t_a^* $, and sends this encrypted information to Bob, who decrypts.

The first step of the synchronization protocol performs a coarse alignment (see algorithm~1 in the supplementary file~\cite{alganju}), where Bob performs a cross-correlation scan between  $\hat{t}_a$ and $\hat{t}_b$  over a range $R<T_{acq}$, using a fixed step size $\delta<R$. This leads to the first estimate of $\tau$, which we refer to as $\tau_{\text{coarse}}$. Note that this coarse alignment is performed only once during initialization (the first round).
Bob then performs a second step commencing in the first round, the fine alignment (see algorithm~2 in the supplementary file~\cite{alganju}).
 Here, the cumulative offset $\tau_{accum}$ is defined as the total relative offset accumulated during previous rounds (set to zero during the first round). 
The fine alignment estimates the relative time offset $\tau$ more precisely, updates Bob's shorter time tag array to $\hat{t}_b$, and updates the cumulative offset $\tau_{accum}$.  Algorithm~2  also automatically detects the time drift $\tau_{drift}$ and eliminates it.

\par Our synchronization protocol is similar to that reported in \cite{pelet2025entanglement}. However, here we incorporate a step-size scan during the first round, making it less vulnerable to large time offsets, such as those possible in Earth-to-satellite channels (target channels for our prototype). 
The synchronization procedure is highly parallelizable (and is implemented as such), making it faster for scanning a wide range of possible time offsets. Our protocol enables real-time implementation, is easy to implement in QKD systems, offers flexibility and robustness in real-world conditions, and is sufficient for entanglement-based key distribution. However, it is the encapsulation of our synchronization protocol within a general QKD-PQC framework that is of main interest here. QKD encryption of the time tag information, although information-theoretic secure, would be expensive due to the large amount of time tag information, likely absorbing more QKD key than produced by the QKD protocol. In ~\cite{dai2020towards}, 128-bit AES encryption is used, with QKD keys only used as a seed as a means to bypass this issue, a process that is not information-theoretic secure. However, our obfuscated encryption of the time tag information allows up to $2^{N_{obs}}$ obfuscated encryption configurations to be used, adding an additional layer of pragmatic security to synchronization.

In addition to the benefits offered to the communications through our enhanced synchronization (\textit{e.g.}, freedom from GPS spoofing and enhanced throughput), the tighter correlation in time of the arriving photons at Alice and Bob provides a layer of enhanced timing-based security. Due to the conservation of energy and momentum, the creation of the photon pairs via the spontaneous parametric down-conversion (SPDC) of a pump photon into a signal and idler pair often results in more temporal correlation than could be expected from separate sources. For example, for conditions similar to our hardware, the work of~\cite{maclean2018direct} has demonstrated a time correlation of order 100~fs (femtoseconds) between photons produced via SPDC. It is also shown that this corresponds to a violation of more than 100 standard deviations from the uncertainty relation that two classical light pulses must satisfy~\cite{maclean2018direct}. This extraordinary coupling of times of arrival of the photons at Alice and Bob leads to an enhanced level of security for our QKD system as an attacker must be able to complete their measurements and resending of the photons in shorter timescales. This enhanced security is in addition to the security already offered in our system through the encryption of Alice's time tagging information referred to earlier.  \textcolor{black}{In the circumstance where the photon pairs remain jointly entangled in polarization and in frequency, additional timing-based security becomes available in our system since, conditioned on both photons being detected by Alice and Bob, the increased spread in energy of the entangled-pair state leads to improved received timing correlations~\cite{Giovannetti2001}.} However, we do point out that although intrinsically embedded in our system, taking advantage of the above enhanced correlations in time would require detector time resolutions beyond those used in our current experimental setup.

\subsection{Hybrid Quantum and Post Quantum Design\label{sec: combined system}}

\begin{figure}[!t]
    \centering
    \includegraphics[width=1\linewidth]{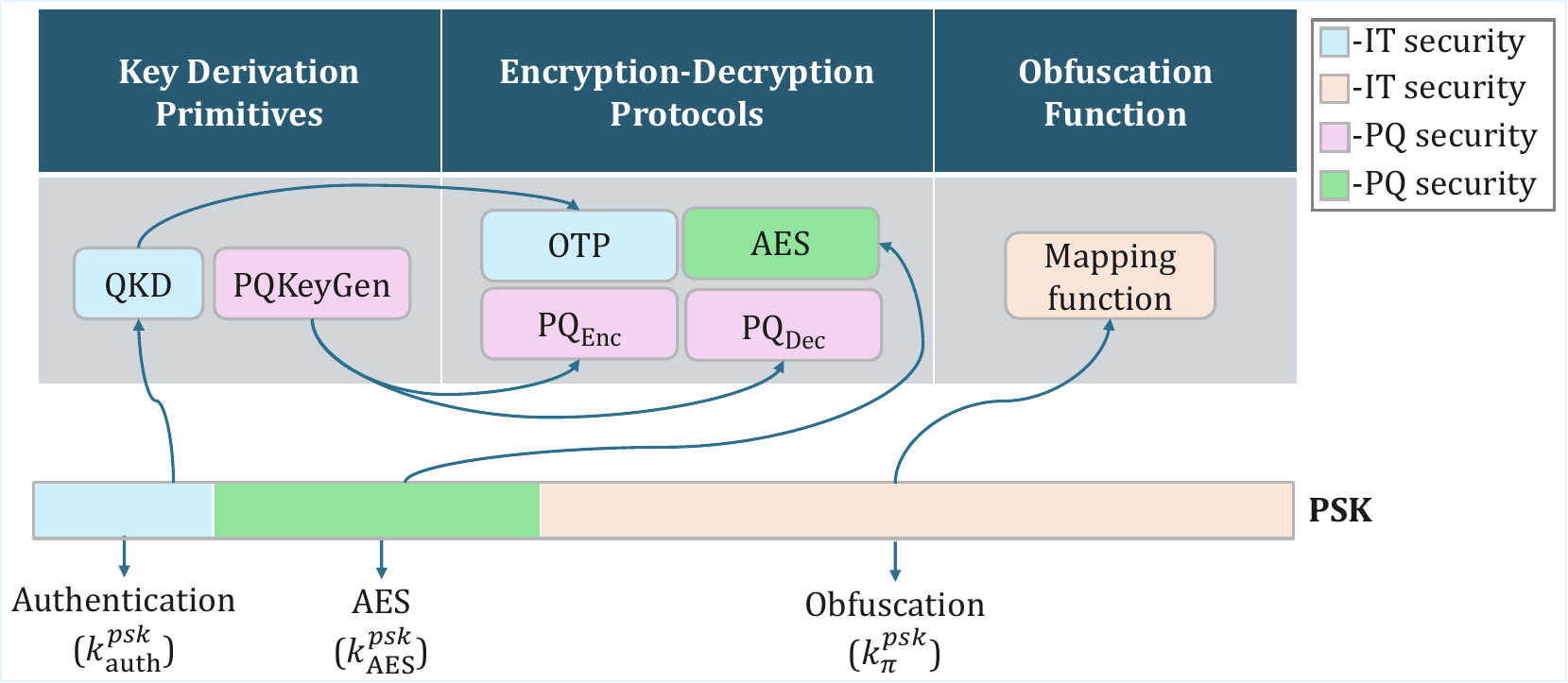}
    \caption{A schematic of the logical flow between the PSK and the different elements of our architecture. The first portion of the PSK is used for QKD authentication. As per normal usage within QKD, when this is exhausted, the new grown QKD key is utilised for ongoing authentication. For AES, the PSK is only used to form a symmetric key. IT refers to information-theoretic security, and PQ refers to quantum-resistant security. $\pi$ denotes the encrypted identifier for an instruction sequence.}
    \label{fig:key_protocol}
\end{figure}

Our architecture comprises QKD and PQC key derivation primitives (KDPs) that generate symmetric QKD keys and asymmetric PQC keys. We also use a subset of a PSK to generate an encrypted sequence of instructions that inform Alice and Bob of the type and order of the encryption-decryption protocols. Additionally, the keys from different KDPs do not combine to generate a hybrid key; instead, a series of encryptions (and their corresponding decryptions) using different keys derived from the KDPs are performed.   
The proposed architecture ensures information-theoretical security in an ideal operational scenario and quantum-resistant security as long as at least one primitive is not compromised. 
An overview of our security architecture is presented in Fig.~\ref{fig:key_protocol}, where the interface between the PSK and the different elements of the architecture is shown. Note that henceforth, when we refer to AES we will mean 256-bit AES, unless otherwise specified.

The length of the PSK, of course, will dictate the complexity of the overall configuration settings. An important parameter is $N_{obs}$, the number of bits available for obfuscation. Regarding the instruction sequence (IS), this will allow for up to $2^{N_{obs}}$ independent IS possibilities. There are several methods for the system to decide on the different forms of the IS. They can be set
\emph{a priori} between Alice and Bob in a secured environment (the same environment in which the PSK is set). However, this does not offer dynamic behaviour (re-adjustment of the configuration settings in the field), instead necessitating a permanent fix of the settings for a pair of devices. In contrast, Bob (the data sender in our system) could encrypt all aspects of the configuration settings in an encrypted message to Alice. This allows for the maximum flexibility and maximum dynamic behaviour, since no configuration settings are pre-set by Alice or Bob (but this will normally allow for less than $2^{N_{obs}}$ configurations to be used). A third possibility is that an \emph{a priori} look-up table could be set and made publicly available where the configuration settings are preset and linked to a unique label (identifier), this label being then encrypted and sent from Bob to Alice. For clarity of exposition, in the following we assume the latter mode of operation to be the case. This chosen operation offers the possibility of dynamic behaviour, albeit from a set of prearranged settings. So that not all the obfuscation key is consumed in one cycle, we set a new length, $\hat{N}_{obs}\ll N_{obs}$, and consume only $\hat{N}_{obs}$ of key bits per cycle (per one use of the look-up table).

\par
\begin{figure*}[!t]
    \centering
    \includegraphics[width=0.84\linewidth]{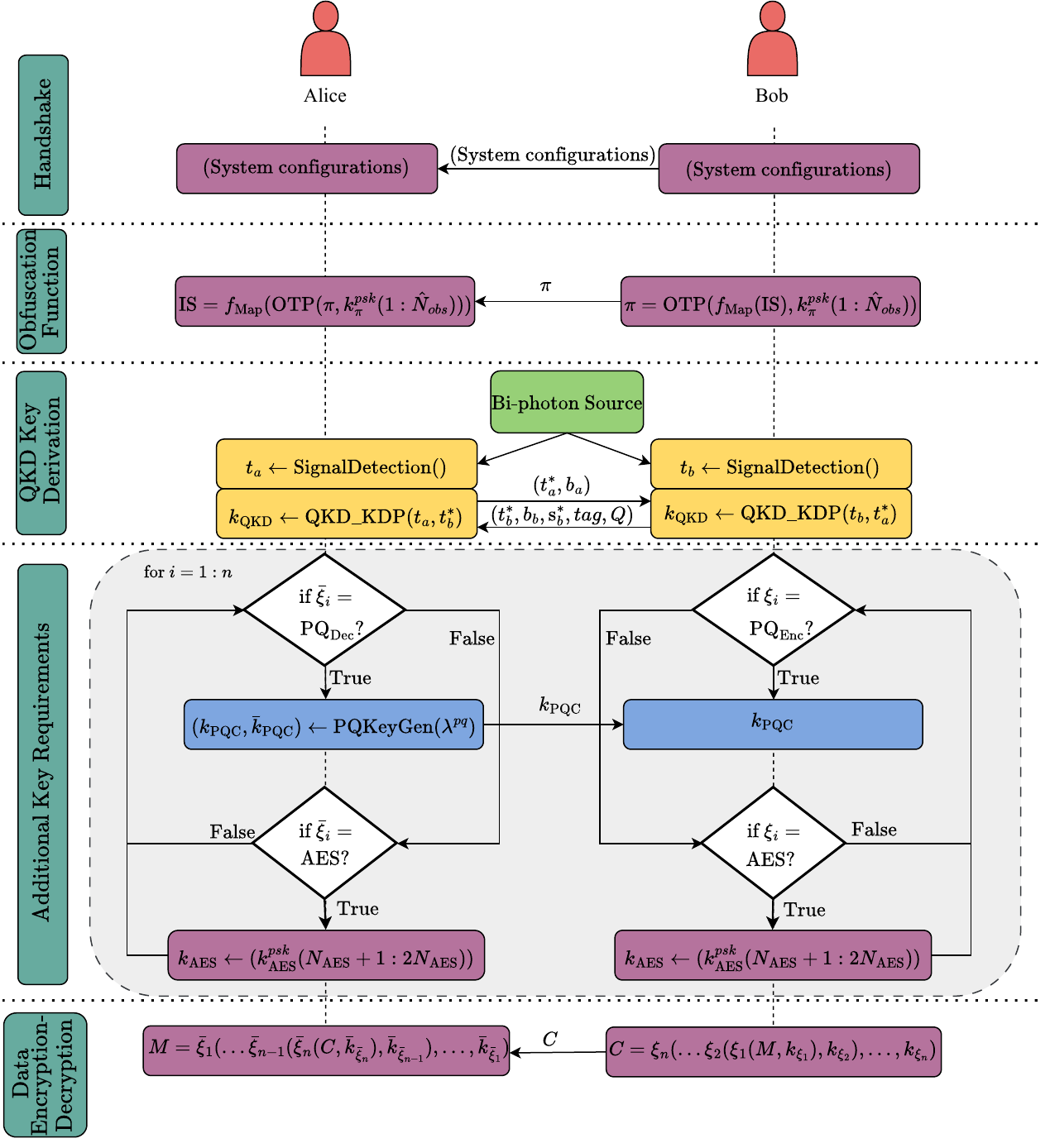}
    \caption{The flowchart describes one cycle of a proposed QKD-PQC system. In the figure, it is assumed that Bob intends to communicate a data message ($M$) to Alice. The term $k^{psk}_\pi(1:\hat N_{obs})$ is used to represent that the next $\hat N_{obs}$ number of unused bits are taken from PSK. Similarly, the term $k^{psk}_{\text{AES}}(N_{\text{AES}}+1:2N_{\text{AES}})$ is used to represent that the next $N_{\text{AES}}$ bits are used for all data encryption-decryption, where the IS contains AES (the first $N_{\text{AES}}$ are used for encryption-decryption in QKD reconciliation). The output of $f_{\text{Map}}(\text{IS})$ is an identifier associated with the instruction sequence that is obtained from the mapping table. The output of Data Encryption obtained via algorithm~3 (the decryption is done with algorithm~4) is the ciphertext denoted by $C$.}
    \label{QKD_PQC_system_flow}
\end{figure*}

  In general, the IS  could contain any information related to the configuration, such as the parameters for any protocol, and instructions on how to encrypt/decrypt  time tag information. However, for clarity, we discuss in detail a specific implementation of the generic framework provided in Fig.~\ref{fig:key_protocol}.
In this specific implementation, we assume that the IS is constructed only from the three underlying primitives (with set parameters) and that the time tag information (and other classical communications associated with QKD reconciliation) are encrypted only using AES.
Our specific implementation, shown in Fig.~\ref{QKD_PQC_system_flow}, possesses two sets of keys. The first set of keys, namely, $k^{psk}_{\text{auth}}$, $k^{psk}_{\text{AES}}$, $k^{psk}_{\pi}$, are used for the authentication in QKD reconciliation, AES encryption of classical data communicated and the derivation of the encrypted identifier of IS, respectively. Note that the encrypted identifier is denoted by $\pi$, and OTP encryption is used to encrypt the identifier.  

The IS can be written as $(\xi_1,\xi_2,\cdots\xi_n)$, where $\xi_i \in \{\text{OTP, AES, PQ}_{\text{Enc}}\}$ and $i = (1,2,\cdots,n)$, where $n$ is the number of encryption-decryption protocols to be applied to the data (the data is the actual message $M$  communicated from Bob to Alice that the entire system is set up to protect). The second set of keys, denoted by $k_{\xi_i}$, is used by the encryption protocols ($\xi_i$), and if the decryption protocol is used then the corresponding decryption keys are represented by $\bar k_{\bar \xi_i}$, where $\bar\xi_i \in \{\text{OTP, AES, PQ}_{\text{Dec}}\}$ are the corresponding decryption protocols. 

Fig.~\ref{QKD_PQC_system_flow} describes the flow chart of the specific protocol, assuming Bob is the sender of the data $M$ while Alice is the receiver of $M$. The steps are summarized below, where one `cycle' of the QKD-PQC setup is the completion of steps (1)-(5) below. 

\subsubsection{\label{handshake}Handshake} Alice and Bob perform an initialization handshake; they share the required system configurations (\textit{e.g.}, security parameters,  FPGA parameters, and post-processing parameters). They also choose $N_{AES}$ bits from $k_{AES}^{psk}$, since in this specific protocol time tags and some other classical communications (\textit{e.g.}, reconciliation information) are always encrypted with AES.

\subsubsection{\label{Seq Der}Obfuscation Function} Bob derives an IS from $2^{\hat{N}_{obs}}$ possible instructions. Next, Bob uses a mapping table $f_{\text{Map}}$ (publicly available) to obtain an identifier for the IS. He encrypts this identifier by XoRing with $\hat N_{obs}$ bits of $k^{psk}_\pi$ to generate $\pi$, which Alice decrypts upon receiving to obtain the same IS.

\subsubsection{\label{QKD key}QKD Key Derivation} Here, the QKD protocol, as described in Sec.~\ref{sec:exp}, is used to establish the QKD keys between Alice and Bob. In Fig.~\ref{QKD_PQC_system_flow}, the process of obtaining the time tag from detecting a quantum signal is denoted by the SignalDetection() function and the post-processing step is denoted by $\text{QKD\_KDP}$. The symbols in the figure corresponding to this step are denoted in Sec.~\ref{sec:exp}.  Note that encrypted time tag arrays are shown as $t_{a}^*,t_b^*$ and encrypted syndrome as $s_b^*$.

\subsubsection{\label{Other key}Additional Key Requirements} This step depends on the IS contents. If the IS contains the element PQ$_{\text{Dec}}$, then Alice generates the asymmetric keys $(k_{\text{PQC}},\bar k_{\text{PQC}}$) using the post-quantum key generation function PQKeyGen~\cite{regev2010learning} with the security strength\footnote{Security strength is a number associated with the amount of work (the number of operations) that is required to break a cryptographic algorithm or system~\cite{barker2020nist}. It is generally a value of the set $\{80, 112, 128, 192, 256,\cdots\}$.} $\lambda^{pq}$ and sends the public key $k_{\text{PQC}}$ to Bob. If the IS contains the AES element, then Alice and Bob obtain the next $N_{\text{AES}}$ bits of $k^{psk}_{\text{AES}}$ to form the key $k_{\text{AES}}$.

\subsubsection{\label{Seq Enc-Dec}Data Encryption-Decryption} At this stage, all the desired keys and instructions are generated. Bob encrypts a data message according to the IS using the algorithm~3 and sends the ciphertext to Alice, which she decrypts using the reverse of the IS via the algorithm~4.
This completes a cycle of our cryptographic architecture (many QKD sessions could reside within one QKD-PQC cycle). 

\subsection{System Advantages}
As mentioned earlier, the encrypted sequence of operations can provide additional burden (thereby providing additional security) on a successful attacker who may have arrived at independent side-channel attacks on QKD and PQC, rendering other hybrid QKD-PQC systems vulnerable. An example scenario would be a successful attack on QKD 
through a power analysis~\cite{powerqkd2},
coupled to a successful power analysis attack on PQC~\cite{attPQC,kamucheka2021power}.  By providing $2^{N_{obs}}$ potential sequences of encryptions (some of which may include a series of multiple encryptions beyond two), currently successful power analysis attacks become ineffective with high probability. A similar argument could be formed based on time analysis attacks on QKD~\cite{taqkd1} and PQC~\cite{ta1pqc1}. In addition, even though both QKD and PQC are safe against ciphertext-only attacks, if other side-channel attacks, such as those mentioned above, are successful, a ciphertext-only attack has a chance of succeeding in other hybrid QKD-PQC systems, but it is ineffective in our system with high probability.

\par Beyond the system model adopted here, the use of an IS for obfuscation can also be applied to different hybrid QKD-PQC configurations, different QKD protocols, different photon sources, different physical channels, and different hardware. The photon source (sources) may be embedded in Alice or (and) Bob. The use of an IS can also be applied outside the post-processing stage. For example, a host of side-channel attacks can be specific to the QKD hardware (or channel) and possibly be relevant in different phases of the QKD protocol (for the BBM92 protocol, these are listed in~\cite{Yin2020}). Extended systems can address such side-channel attacks through various means, such as setting intermittent use of the quantum channels for purposes beyond QKD or setting detector parameters, with the information necessary to coordinate these settings being secretly communicated as part of the IS. 
We do not claim that our approach is the only method to counter hardware-based side-channel attacks: rather, we offer a more flexible and easy-to-implement solution.

\subsection{Formal Security Considerations}
\textcolor{black}{
As stated above, our new system presents the adversary with additional challenges in terms of the resources needed to attack due to the exponential number (in $\hat N_{obs}$) of possible configurations presented to her. In the context of a successful attack on the QKD primitive (assumed throughout this subsection), our system provides additional protection against a successful side-channel attack on the PQC primitive within a single QKD-PQC hybrid. However, this is not a formal PQC security. In the context of classical PQC security frameworks\footnote{ By definition, there is no formal security associated with PQC which is the equivalent of the information-theoretic security provided to QKD. Rather, there is an attempt to map algorithms to known hard problems and argue that no attack is possible (known) within a certain number of operations. The NIST process explicitly characterizes five categories of security definitions each associated with a different number of operations that a brute-force attack would need to find the key~\cite{moody2021nist}.} the issue of leakage caused by side-channel attacks, even on a single use of a PQC primitive remains an open problem in terms of security proofs~\cite{hoffmann2023polka}.  However, it is straightforward to show that if many different PQC primitives are deployed as in our system, then only one primitive needs to deliver a secure key for the entire chain to deliver a secure key. This is so because in the chain of XoR operations on sequential key outputs $k_1 \oplus k_2\dots k_n$, only one of the $k_i$'s needs to be secure for the entire sequence of operations to be secure. The more difficult question is what can we say about PQC security in the case where a partial leakage occurs in two or more of the PQC primitives? An answer to this will likely have to wait for proof of PQC security to be given to the single-primitive leakage case.  However, we hope that the present work inspires further theoretical development in such formal analyses. Pending these new developments, we must rely on one of the following; the exponential use of resources, the scenario where at least one PQC primitive in a chain of $n$ such primitives remains secure, or the intuitive understanding that a series of partial leakages should provide more security than a single leakage in scenarios where the encrypted message from each PQC use remains hidden from the adversary.}

\textcolor{black}{
Beyond these general statements on security, it may also be useful to mention three more specific threat models,  highlighted in previous works, which provide poor security outcomes but are compensated for to some extent in our system. (i) Consider the case of simultaneous side-channel attacks that reveal all secret keys from the PQC key generation primitives, clearly a bad outcome for most systems. However, in our system, even in this bad scenario, Eve still requires $\mathcal{O}(2^{(\hat{N}_{obs}/4)(N_{AES}+2)})$ operations to compromise the message. (ii) In the context of cascade encryptions, some works have pointed out a counter-intuitive issue~\cite{cryptography7040049, rogaway2018onion,gavzi2009cascade,10.1007/978-3-642-38348-9_25}. This is  when comparing cascaded encryption of $l$ encryptions each with $x$-bit keys, and a single encryption with an $lx$-bit key, the number of operations that Eve needs to perform a brute-force attack of the latter is $\mathcal{O}(2^{lx})$, while in the former it can be $\mathcal{O}(2^{(lx/2+1)})$~\cite{10.1007/978-3-642-38348-9_25}. The reason for this counter-intuitive outcome can be traced back to a "meet-in-the-middle attack"~\cite{gavzi2009cascade}. (iii) Another issue within the multiple encryption scenario is that which occurs under a "re-encryption" attack~\cite{cryptography7040049}. In one manifestation of this attack, Eve attempts to convince the receiver to accept multiple copies of the same message from the transmitter. Counter intuitively to some, it can be shown that by decrypting and re-encrypting the final encryption with a public key no advantage of the multiple encryptions in protecting against this attack is forthcoming. It is straightforward to understand that the counter-intuitive results in (ii) and (iii) are both negated in our system since Eve has no access to the encryption-decryption oracle, as it is an information-theoretically secure $2^{\hat{N}_{obs}}$ permutations of different encryption algorithms with independent PSK keys.
}

\section{\label{sec:results}Results and Discussion}
We implemented a fully integrated real-time prototype QKD-PQC system in laboratory settings. The time synchronization for QKD is done by performing a cross-correlation analysis between the detection events at the different parties receiving entangled photons.
The QKD experiment is conducted over a channel length of 1.5~m. An experimental loss of 10.3~dB is measured for the signal, accounting for transmission, optical, and detection losses. The EPS (QES2, Qubitekk) provides tunability over power and temperature, and can be configured for both degenerate and non-degenerate cases. In our setup, the source is configured for the degenerate case producing a degenerate signal and idler pair at 810~nm.
The power of the source is varied and visibility is calculated for different power values to optimize the source for maximum visibility and count rate (see results in~\cite{rani2025combined}). In our case, the source is set at a power value of milliwatts, achieving a coincidence rate of 10,000~s$^{-1}$ (just after the source, measured in fiber). The visibility of the source at the set power value is measured as $0.873\pm0.019$. The detection efficiency of the detectors is 65\% (Excelitas, SPCM-AQRH-14) with a background count of 100~s$^{-1}$. The time tagging device is the OPX+ from Quantum Machines, which also serves as our post-processing unit and consists of an onboard FPGA with a timing resolution of 1~ns. \textcolor{black}{For QKD reconciliation, a low density parity check  matrix  is adopted based on the progressive edge growth algorithm in~\cite{xiao2004improved} with a code rate $R_c=0.5$, and degree distributions optimized for QKD as described in~\cite{elkouss2009efficient}.} The length of $N_{raw}$ is set to $40,000$. Both the post-processing and the PQC module are handled in personal computers with 16~GB RAM and a 2.9~GHz clock speed GPU for both Alice and Bob. The PQC key generation and encryption-decryption modules are taken from the Post Quantum Safe Lib module~\cite{pqc_git}, and the post-processing module was developed in MATLAB v2021. 
\par
\begin{figure}[ht]
		\centering		
        \includegraphics[width=1\linewidth]{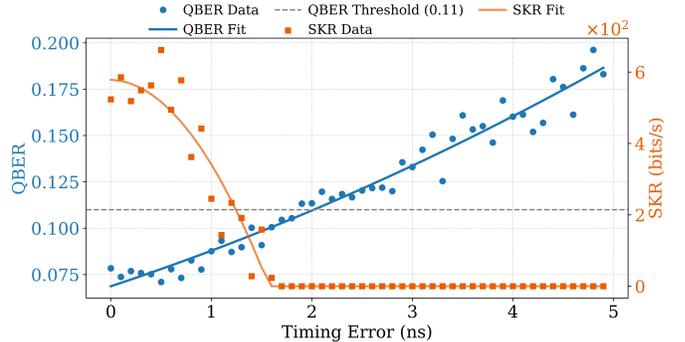}
		\caption{\textcolor{black}{QBER and SKR as functions of timing error. Due to imperfect reconciliation, SKR drops to zero as QBER exceeds 0.1 (not 0.11). Here, the timing error is the standard deviation of the timing jitter in the photon detectors (Gaussian noise assumed). The coincidence window used to determine the counts is $1$~ns. Note, these key rates are upper limits  due to the value of $N$ adopted. Approximately a factor of ten more collection time is required for the rates shown to be achievable.}}
        \label{fig:QCSg1}
\end{figure}

 We examined the impact of timing errors (introduced assuming Gaussian jitter noise in Bob's time tag array $t_b$) on the performance of the QKD system in terms of the secret key rate~(SKR), which is calculated using our key rate equation. 
Fig.~\ref{fig:QCSg1} illustrates the critical role played by precise time alignment in optimizing QKD performance. An increased timing error reduces the coincidence detection rate, which in turn increases the QBER and lowers the SKR. 

\par
\begin{figure}[ht]
		\centering		
        \includegraphics[width=1\linewidth]{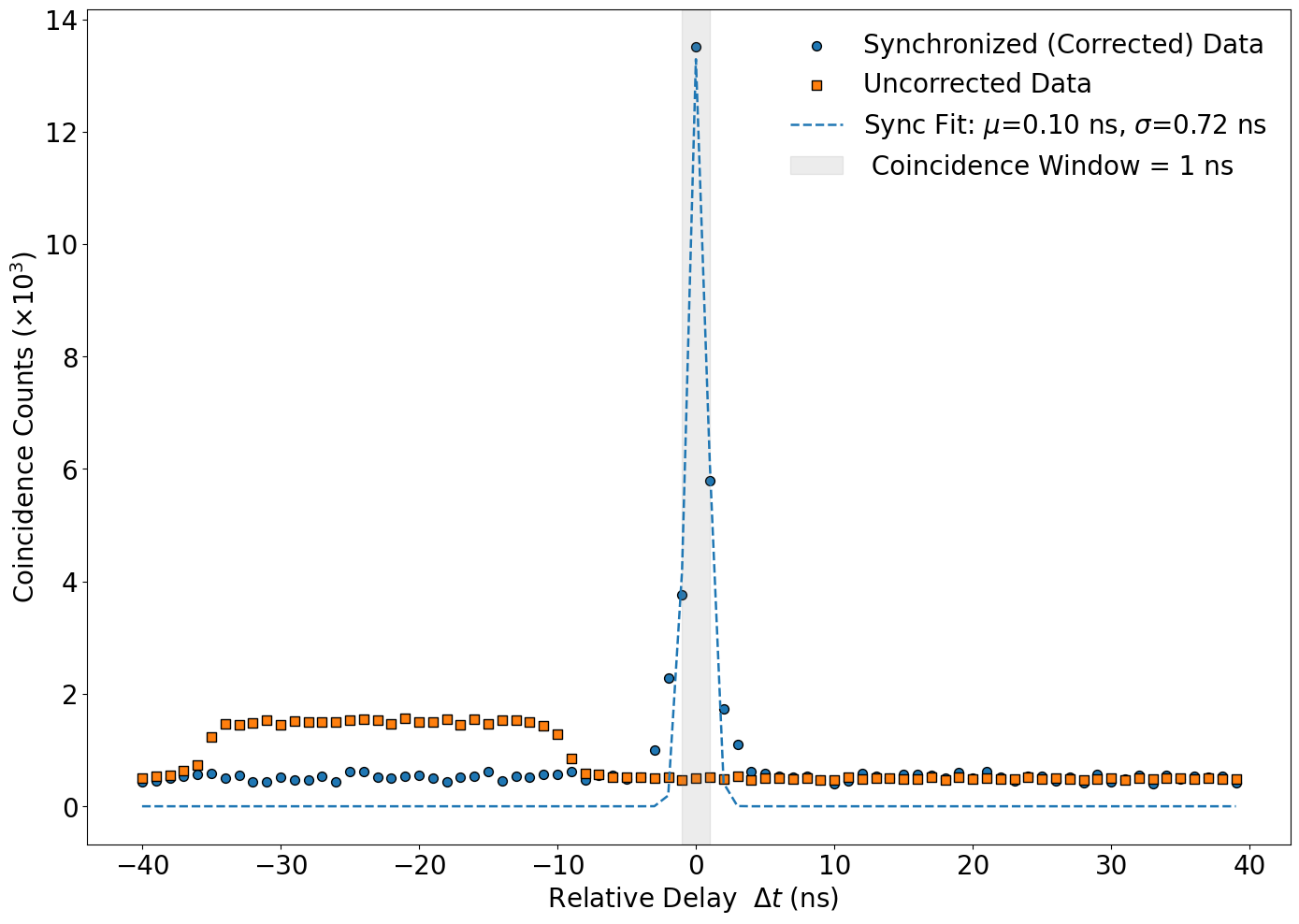}
		\caption{\textcolor{black}{Coincidence counts as a function of a relative delay, $\Delta t$, added to Bob's time tag information. Counts are shown before (orange) and after (blue) application of the synchronization protocol.  For each $\Delta t$, coincidences are counted only when Alice’s and Bob’s time tags match within a 0.5~ns window. The input information used here had a time offset of 10~ns and a timing drift of 1~ns per 250~ms applied. The points representing the information after applying our synchronization show a centered coincidence peak at $\Delta t = 0$,  indicating the time offset and drift have been successfully removed. The data shown is accumulated over 4.5~s.}}
        \label{fig:reldrift}
\end{figure}

 In order to test our synchronization protocol (see appendix), we introduce a 10~ns (nanoseconds) relative time offset $\tau$ and a time drift $\tau_{drift}$ of 1~ns per 250~ms (milliseconds) in Bob's time tags. The parameter values used in the protocol are detailed in Table~\ref{tab:synch_parameters}.       
On average, the protocol takes approximately $5~\mathrm {ms}$ per round, except during the first round, where the coarse alignment step takes approximately 0.5~sec due to heavier processing. 
In Fig.~\ref{fig:reldrift}, which highlights the overall performance of the synchronization protocol, the abscissa axis represents a relative delay added to the input time-tag information in order to search for a peak in the correlation between Alice and Bob's input time tag arrays. If there were no synchronization errors in the input information, a peak should be found at zero. However, as we can see, without the application of our synchronization, no coincidence peak is evident in the counts at zero relative delay between Alice and Bob (as shown by the orange scatter points in Fig.~\ref{fig:reldrift}). Instead, a broad distribution of low coincidence counts is observed in the relative delay range (the increase in counts between $-10~\text{ns to}-38~\text{ns}$ can be attributed to a partial success in removing the drift component). If such a time tag array were used in QKD without correction, the QKD protocol would fail. Upon applying our synchronization, we remove the relative time offset and the time drift from the time tag array and retrieve a sharp coincidence peak at zero relative delay (as shown by the blue scatter points in Fig.~\ref{fig:reldrift}) thereby illustrating the success of our protocol. From Fig.~\ref{fig:reldrift}, we also observe that a coincidence window of 1~ns would be an excellent choice for QKD implementation. 

\begin{table}[h]
\centering
\caption{Synchronization Parameters }
\begin{tabular}{ll}
\toprule
\textbf{Parameter} & \textbf{Value} \\
\midrule
Acquisition  period ($T_{acq}$) & 50~ms \\
Coarse scan step size (\( \delta \)) & 200~ns \\
Coincidence peak width (FWHM) & 1~ns \\
Coarse scan range (\( R \)) & \(\pm1\)~ms \\
Fine alignment bin width (\( w \)) & 1~ns \\

\bottomrule
\end{tabular}
\label{tab:synch_parameters}
\end{table}

\begin{table*}[h!]
    \centering
    \caption{The mean execution times (in seconds) of a standard QKD system with OTP encryption on both Alice's side and Bob's side, compared with the mean execution time of our combined QKD-PQC system for the same message and same system settings on both Alice's side and Bob's side. The Classical Communications is the total time taken in the classical communication of information in the QKD post-processing. The synchronization only occurs in Bob's side; hence, no execution time for synchronization in Alice's side. The codeblock `Message Enc-Dec' in a standard QKD system corresponds to a simple XoR operation, while in our hybrid system it is the \textcolor{black}{Data Encryption-Decryption} defined in step (\ref{Seq Enc-Dec}) of Sec.~\ref{sec: combined system}.}
    \label{tab:exe_QKD-PQC}
    \begin{tabular}{l c c c c c}
        \toprule
        \multirow{2}{*}{Codeblock} & \multicolumn{2}{c}{Standard QKD system with OTP (Time in sec.)} & \multicolumn{2}{c}{Our QKD-PQC system (Time in sec.)}\\
        \cmidrule(lr){2-3} \cmidrule(lr){4-5}
        & Alice & Bob & Alice & Bob \\
        \midrule
        Synchronization & -- & $0.54\pm0.08$ & -- & $0.58\pm0.18$  \\
        Time Tag Filtering & $0.0004\pm0.00006$ & $0.0041\pm0.0014$ & $11.84\pm0.23$ & $22.82\pm0.63$ \\
        Sifting & $0.0004\pm0.0003$ & $0.0004\pm0.0005$ & $0.26\pm0.01$ & $0.40\pm0.03$\\
        QBER Estimation & $0.0002\pm0.0002$ & $0.0003\pm0.0002$ & $0.0002\pm0.0001$ & $0.0002\pm0.0001$ \\
        Error Correction & $0.12\pm0.24$ & $0.33\pm0.41$ & $0.19\pm0.25$ & $0.37\pm0.27$ \\
        Privacy Amplification & $0.20\pm0.01$ & $0.21\pm0.01$ & $0.21\pm0.01$ & $0.21\pm0.01$\\ 
        Obfuscation & -- & -- & $0.0003\pm0.0003$ & $0.0005\pm0.0008$\\
        PQC Key Share & -- & -- & $0.95\pm0.011$& $1.29\pm0.245$\\
        Message Enc-Dec & $0.002\pm0.003$ & $0.006\pm0.009$ & $0.966\pm0.084$& $2.24\pm0.070$\\
        Classical Communications & $14.46\pm1.15$& $7.74\pm0.39$ & $32.91\pm1.13$& $16.11\pm0.39$\\
        \midrule 
        Total execution time (in sec.) & $14.82\pm1.17$ & $8.82\pm0.57$ & $47.32\pm1.18$ & $44.03\pm0.85$ \\
        \bottomrule
    \end{tabular}
\end{table*}

\begin{figure}[h]
    \centering
    \includegraphics[width=01\linewidth]{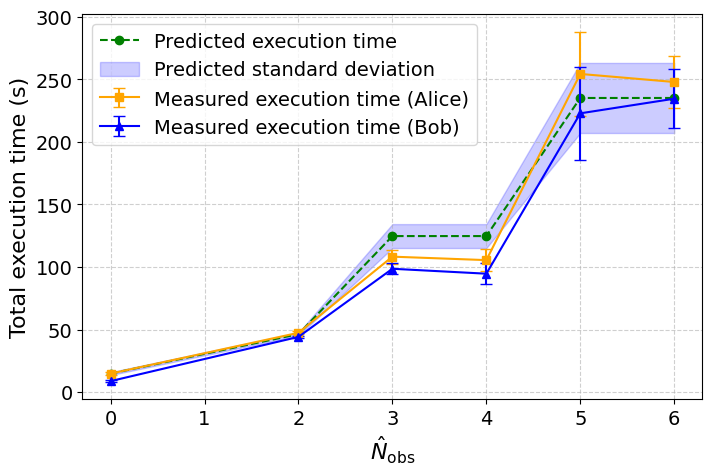}
\caption{The scaling of the total execution time of our system with respect to increasing $\hat{N}_{obs}$ values. The predicted line linearly scales as $\lceil n/2\rceil$ (also with the changing data size that arises from consecutive encryptions). The contribution from each codeblock referred in the first column of  Table~\ref{tab:exe_QKD-PQC} is summed to obtain the shown total execution time.}
    \label{fig:n_six}
\end{figure}

In addition, we compared the performance and overhead of the combined QKD-PQC system.  Table~\ref{tab:exe_QKD-PQC} compares the execution times, in seconds, of a standard (stand-alone) QKD system with OTP encryption with our new hybrid QKD-PQC system (with the same settings for the QKD part). To demonstrate the hybrid design in an experimental setting, we first used the mapping table $f_{\text{Map}}$, where the four possibilities of  identifier $\rightarrow$ IS are:~$00 \rightarrow (\text{OTP, AES})$, $01 \rightarrow (\text{AES, OTP})$, $10 \rightarrow (\text{OTP, PQ}_\text{Enc})$ and~$11 \rightarrow (\text{PQ}_\text{Enc}, \text{OTP})$.
This determines $n=2$ and $\hat N_{obs} =2$, which sets the number of instruction sequences to be~$2^{\hat N_{obs}} = 4$, resulting in the consumption of~2 bits of PSK per cycle to communicate one of the four configuration choices. For AES, we set the symmetric key, $N_{\text{AES}} =256$~bits.
The execution times for each codeblock were recorded over 10 runs and the mean and standard deviations were computed. A stable \textcolor{black}{SKR of $580\pm59$ bits/s} was observed and a QBER of $0.0745\pm0.0047$. The increase in time in the codeblocks from `Synchronization' to `Error Correction' is due to the fact that we perform an additional AES encryption-decryption of classical information in the post-processing step. As can be seen, by the comparison of total times, there is approximately 3.19 times overhead on Alice's side and  4.99 times overhead on Bob's side in execution times in our combined QKD-PQC system compared to a standard QKD with OTP encryption. Assuming that a mapping table $f_{\text{Map}}$ is constructed such that, on average asymptotically, an IS contains $n/2$ encryption-decryptions with OTP, $n/2$ with AES and $n/2$ with PQC, the execution time for that IS will increase by approximately a factor of $n/2$. \textcolor{black}{ This can be seen better from Fig.~\ref{fig:n_six} where additional PQC is sequences (up to 6) are included in the calculation. For simplicity in this figure, we have adopted $n=\hat{N}_{obs}$ for all experiments. We will not breakdown the detailed subset of times within the total execution time for these additional experiments as we have done in Table~\ref{tab:exe_QKD-PQC} for $n=\hat{N}_{obs}=2$, suffice to say that similar scaled trends were observed for these additional times. Without considering any optimization techniques, such as parallelization of computations, Fig.~\ref{fig:n_six} demonstrates that large values of $n$ and $\hat{N}_{obs}$ can be executed on standard computing platforms on timescales smaller than the time to collect the necessary quantum signals for QKD delivery (in most scenarios).}

There are many variations and extensions on the actual implementation of the obfuscated hybrid QKD-PQC we have reported here. For example, the use of lossy compression codes (\textit{e.g.}, Universal and Huffman coding) could be invoked to optimize the per bit of PSK used in the obfuscation function.  Other modules or parameters that undergo obfuscation could be introduced. Examples of these would be the channel or reconciliation codes used in classical communications and the block length of the codes. These could be invoked through more sophisticated handshake protocol modules. Other protocols could be merged into our system, such as quantum location verification, leading to position-dependent cryptography  (\textit{e.g.,} see~\cite{mal16a, mal16b}. The merging of security attributes such as those provided by other hybrid QKD-PQC systems (see Introduction) could be convolved within our system, the
anonymous routing of~\cite{otero2025onion} being an example. The full use of photon time correlations, in SPDC creation, for enhanced timing-based security~\cite{maclean2018direct} is another avenue for future work.
\textcolor{black}{Finally, we have not shown that all possible instruction sequences allowed by our system lead to data transfers that are formally quantum resistant to the same dual side-channel attacks that can make previous QKD-PQC architectures insecure. Formal analyses in this direction would be useful.}

\section{\label{sec:conclusions}Conclusion}
We proposed a new hybrid cryptographic system that generates QKD and PQC keys whilst obfuscating in an information-theoretic manner both the sequence on how these keys are used and relevant system parameters. We achieved this, in part, by using an encrypted instruction sequence that indicated a specific order of encryption-decryption of data messages. Our system also provided a GPS-free time-synchronization feature within the QKD key generation system. The described system reduces the vulnerability of hybrid QKD-PQC solutions to known (and yet-to-be discovered) simultaneous practical attacks on the underlying cryptographic primitives.
We carried out a full experimental deployment, comparing the time overhead of our hybrid system with a standard QKD system, demonstrating the feasibility of the former. Our new architecture illustrates the important role that hybrid systems can play in real-world deployments of the QKD and PQC primitives. Future users of our hybrid QKD-PQC system who remain skeptical about QKD security issues in a practical setting can be assured that the worst-case security outcome for them is enhanced PQC security relative to the level of PQC security they normally accept.  Regarding QKD itself, we believe that ultimately all QKD deployments will be deployed in a hybrid fashion similar to that reported here.

\section*{\label{sec:Acknowledgments}Acknowledgments}
The authors thank Dr. Dushy Tissainayagam from Northrop Grumman Australia for constructive feedback and support received during the course of this experiment. This research has been carried out as a project co-funded by Northrop Grumman Australia, and the Defence Trailblazer Program, a collaborative partnership between the University of Adelaide and the University of New South Wales co-funded by the Australian Government, Department of Education.

\renewcommand*{\bibfont}{\normalfont\small}
\singlespacing
\bibliographystyle{IEEEtran}
\bibliography{IEEEabrv,bibfile}

\newpage
\clearpage
\color{black}
\noindent
\textbf{Supplementary File: Algorithms for ``Obfuscated Quantum and Post-Quantum Cryptography,'' by Rani \textit{et al} 2025.}

For completeness, here we list the algorithms used in the specific implementation presented in this work. 

The GPS-free synchronization protocol described in Section~\ref{sec:timetag} uses algorithms~\ref{coarse_algo} and~\ref{alg:multi-interval-sync} below. Note that most of the synchronization processing uses the latter algorithm, as the former is used only in the first round.

\begin{algorithm}
\caption{Coarse alignment Step (only for $1^{st}$ round)}\label{coarse_algo}
\label{alg:multi-interval-sync}
\SetAlgoLined
\DontPrintSemicolon
\SetKwComment{Comment}{/* }{ */}

\KwIn{The time tag array of Alice $\hat{t}_a$:  the time tag array of Bob $\hat{t}_b$: the step size $\delta$:  the coincidence peak width  of the EPS source FWHM:  the coarse scan range $R$. }
\KwOut{Initial time offset $\tau_{coarse}$}

$W \gets \delta + 2\,\text{FWHM}$ \CommentSty{/* Coincidence window width */}\;
$\mathcal{S} \gets \{-R, -R+\delta, \dots, R\}$ \CommentSty{/* Time shifts to scan over */}\;
\For{$\nu \in \mathcal{S}$} 
{  \CommentSty{/* Using MATLAB’s Parallel Computing Toolbox here*/}

  $\hat{t}_b' \gets \hat{t}_b + \nu$\;
  $C[\nu] \gets |\{(a, b') : a \in \hat{t}_a, b' \in \hat{t}_b', |a - b'| \leq W\}|$\;
  \CommentSty{/* Coincidence count at $\nu$ */}
}
$\tau_{\text{coarse}} \gets \arg\max_\nu C[\nu]$\;

{\tiny \CommentSty{Note: The window $W$ ensures that true coincidence events (entangled photon pair coincidence events) near the edges of adjacent steps are not missed.}}
\end{algorithm}

\begin{algorithm}
\caption{Fine alignment Step}\label{fine_algo}
\label{alg:multi-interval-sync}
\SetAlgoLined
\DontPrintSemicolon
\SetKwComment{Comment}{/* }{ */}

\KwIn{The time tag array of Alice $\hat{t}_a$:  the time tag array of Bob $\hat{t}_b$: the initial time offset $\tau_{coarse}$:  the bin width $w$:   the cumulative offset $\tau_{\text{accum}}$. }
\KwOut{ The estimated time offset $\tau$: the updated time tag array $\hat{t}_b$: $\tau_{\text{accum}} $}
$\hat{t}_b \gets \hat{t}_b + \tau_{\text{coarse}}+\tau_{\text{accum}} $\;
$(\hat{t}_a^{\dagger}, \hat{t}_b^{\dagger}) \gets \{(a, b) : a \in \hat{t}_a, b \in \hat{t}_b^{\dagger}, |a - b| \leq W\}$\;
\For{$j = 1 : |\hat{t}_a^{\dagger}|$}{
  $\Delta \hat{t}_j \gets \hat{t}_a^{\dagger}[j] - \hat{t}_b^{\dagger}[j]$\;
  $b_j \gets \text{bin}(\Delta \hat{t}_j)$\;
}
$b^{\dagger} \gets \arg\max_b |\{j : b_j = b\}|$\;
$\tau_{\text{fine}} \gets b^{\dagger} \cdot w$\;

\CommentSty{/* Compute total time offset, update cumulative offset, update Bob’s time tags */}

$\tau \gets \tau_{\text{coarse}} + \tau_{\text{fine}}$\;
$\hat{t}_b\gets \hat{t}_b + \tau$\;
$\tau_{\text{accum}} \gets \tau_{\text{accum}} + \tau$\;
$\tau_{\text{coarse}} \gets  0$\;
{\tiny \CommentSty{Note: Other than the first round, $\tau_{drift} $ is estimated by the $\tau$ (output of algorithm).}}
\end{algorithm}

The specific implementation of the hybrid QKD-PQC protocol described in Section~\ref{sec: combined system} uses algorithms~\ref{aman1} and ~\ref{aman2} below. These algorithms are implemented in the last step of the control flow shown in Fig.~\ref{QKD_PQC_system_flow}, and largely relate to the encryption and decryption of the data $M$ using the keys created by the hybrid system.

\begin{algorithm}[hbt!]
\caption{\textcolor{black}{Data Encryption}}\label{Seq_enc_algo}
\label{aman1}
\SetAlgoLined
\DontPrintSemicolon
\SetAlgoNoLine
\SetKwComment{Comment}{/* }{ */}
\KwIn{$(\text{Map}, M, \xi, k_\xi,k_{psk})$}
\KwOut{$C$}

\CommentSty{/*Step 1: Bob generates an instruction sequence using a mapping table based on a pre-shared key.*/}\\
$\pi \gets f^\xi_\mathrm{Map}(k_{psk})$ \CommentSty{ /* $\pi = (\xi_{(1)},\xi_{(2)},\cdots,\xi_{(n)})$*/} \\ 
\CommentSty{/* Step 2: Bob generates the key list based on the sequence $\pi$.*/}\\
$k_\pi \gets \mathrm{SeqEncKeys}(\pi, k_\xi)$;
$C_{temp} \gets M$\\
\For{$i\gets1$ \KwTo $\text{len}(\pi)$}
{
$C_{temp} \gets \xi_{\pi(i)}(C_{temp},k_{\pi(i)})$
}
$C \gets C_{temp}$\\
\CommentSty{/* Step 3: Bob sends ciphertext $C$ to Alice.*/}
\end{algorithm}

\begin{algorithm}[hbt!]
\caption{\textcolor{black}{Data Decryption}}\label{Seq_dec_algo}
\label{aman2}
\SetAlgoLined
\DontPrintSemicolon
\SetKwComment{Comment}{/* }{ */}
\KwIn{$(\text{Map}, M, \bar{\xi}, \bar{k}_{\bar{\xi}},k_{psk})$}
\KwOut{$M$}
\CommentSty{/*Step 1: Alice, upon receiving ciphertext $C$ initiates the decryption process.*/}\\
$\pi^{-1} \gets \pi \gets f_{\text{Map}}^{\bar{\xi}}(k_{psk})$\\ 
\CommentSty{/*Step 2: Alice generates the key list based on the sequence $\pi^{-1}$.*/}\\
$\bar{k}_\pi \gets \mathrm{SeqDecKeys}(\pi^{-1}, \bar{k}_{\bar{\xi}})$;
$M_{temp} \gets C$\\
\For{$i\gets1$ \KwTo $\text{len}(\pi)$}
{
$M_{temp} \gets \bar{\xi}_{\pi(n-i-1)}(M_{temp},\bar{k}_{\pi^{-1}(i)})$
}
$M \gets M_{temp}$\\

\end{algorithm}

\end{document}